\documentclass[aip,rsi,twocolumn,amsmath,amssymb, reprint]{revtex4-1}
\usepackage[pdftex]{graphicx}
\usepackage{hyperref}

\begin{document}


\title{Vibration-induced electrical noise in a cryogen-free dilution refrigerator: characterization, mitigation, and impact on qubit coherence}



\author{Rachpon Kalra}
\altaffiliation[Now at ]{School of Mathematics and Physics, University of Queensland, Brisbane QLD 4072, Australia}
\author{Arne Laucht}
\author{Juan Pablo Dehollain}
\altaffiliation[Now at ]{QuTech \& Kavli Institute of Nanoscience, TU Delft, 2628CJ Delft, The Netherlands}
\author{Daniel Bar}
\author{Solomon Freer}
\author{Stephanie Simmons}
\altaffiliation[Now at ]{Physics Department, Simon Fraser University, British Columbia, Canada}
\author{Juha T. Muhonen}
\altaffiliation[Now at ]{Center for Nanophotonics, FOM Institute AMOLF, Science Park 104, 1098 XG, Amsterdam, The Netherlands}
\author{Andrea Morello}
\email{a.morello@unsw.edu.au}
\affiliation{Centre for Quantum Computation and Communication Technology, School of Electrical Engineering and Telecommunications, UNSW Australia, Sydney NSW 2052, Australia}

\date{\today}

\begin{abstract}
	Cryogen-free low-temperature setups are becoming more prominent in experimental science due to their convenience and reliability, and concern about the increasing scarcity of helium as a natural resource. Despite not having any moving parts at the cold end, pulse tube cryocoolers introduce vibrations that can be detrimental to the experiments. We characterize the coupling of these vibrations to the electrical signal observed on cables installed in a cryogen-free dilution refrigerator. The dominant electrical noise is in the 5 to 10~kHz range and its magnitude is found to be strongly temperature dependent. We test the performance of different cables designed to diagnose and tackle the noise, and find triboelectrics to be the dominant mechanism coupling the vibrations to the electrical signal. Flattening a semi-rigid cable or jacketing a flexible cable in order to restrict movement within the cable, successfully reduces the noise level by over an order of magnitude. Furthermore, we characterize the effect of the pulse tube vibrations on an electron spin qubit device in this setup. Coherence measurements are used to map out the spectrum of the noise experienced by the qubit, revealing spectral components matching the spectral signature of the pulse tube. 
\end{abstract}


\maketitle 


\section{Introduction}

Traditional `wet' dilution refrigerators rely on liquefied helium, which is a limited natural resource that is becoming increasingly scarce and, therefore, expensive. While a system can be installed for reliquefaction, the necessary equipment and infrastructure is quite expensive and, even then, $\sim 10$\% losses are common. Further disadvantages of wet fridges include the small size of the sample space, limited by the narrow-neck dewars required to minimize helium boil-off, and the need to refill the helium dewar every 2-4~days which may interrupt sensitive experiments. These factors have led to the increasing popularity of cryogen-free systems, where a pulse tube (PT) cooler is used to cool the dilution unit to below 4~K. These `dry fridges' do, however, have major sources of vibrational and acoustic noise \cite{Tomaru2004,Riabzev2009}. The necessary compressor and rotary valve, used in modern systems, are sources of tangible vibrations external to the cryostat. Inside the pulse tube, the dynamics of the helium gas can produce acoustic vibrations that are directly coupled to the cold head.

Significant efforts have been made by manufacturers and researchers \cite{Baer2014phd,Tian2012rsi} to minimize the coupling of these vibrations to the sample-space in the cryostat, as will be described in Section~\ref{sec:Setup}. For example, atomic scale microscopy in cryogen-free systems required further in-house modifications to achieve extreme mechanical isolation \cite{Pellicione2013,denHaan2014}. In the context of qubit experiments, mechanical vibrations can contribute to dephasing through motion in an inhomogeneous magnetic field, or resulting electrical noise on control lines (Sections~\ref{sec:CableComparison} and \ref{sec:coherence}).

In this paper, we report on how vibrational noise from the pulse tube significantly couples into the electrical signal measured on our cables, over a bandwidth of up to 40~kHz. We characterize this noise and, by testing the performance of different types of cables, learn that the dominant mechanism through which the vibrations couple into the electrical signal is via triboelectric effects \cite{Ong1987}. We then use the cables that yield the minimum amount of voltage and current noise to test the coherence of an electron spin qubit. Performing noise spectroscopy clearly reveals that the vibrations from the pulse tube translate into noise that contributes to qubit decoherence.

\section{Measurement setup}
\label{sec:Setup}

\begin{figure}[t]
	\includegraphics[width=\columnwidth, keepaspectratio = true]{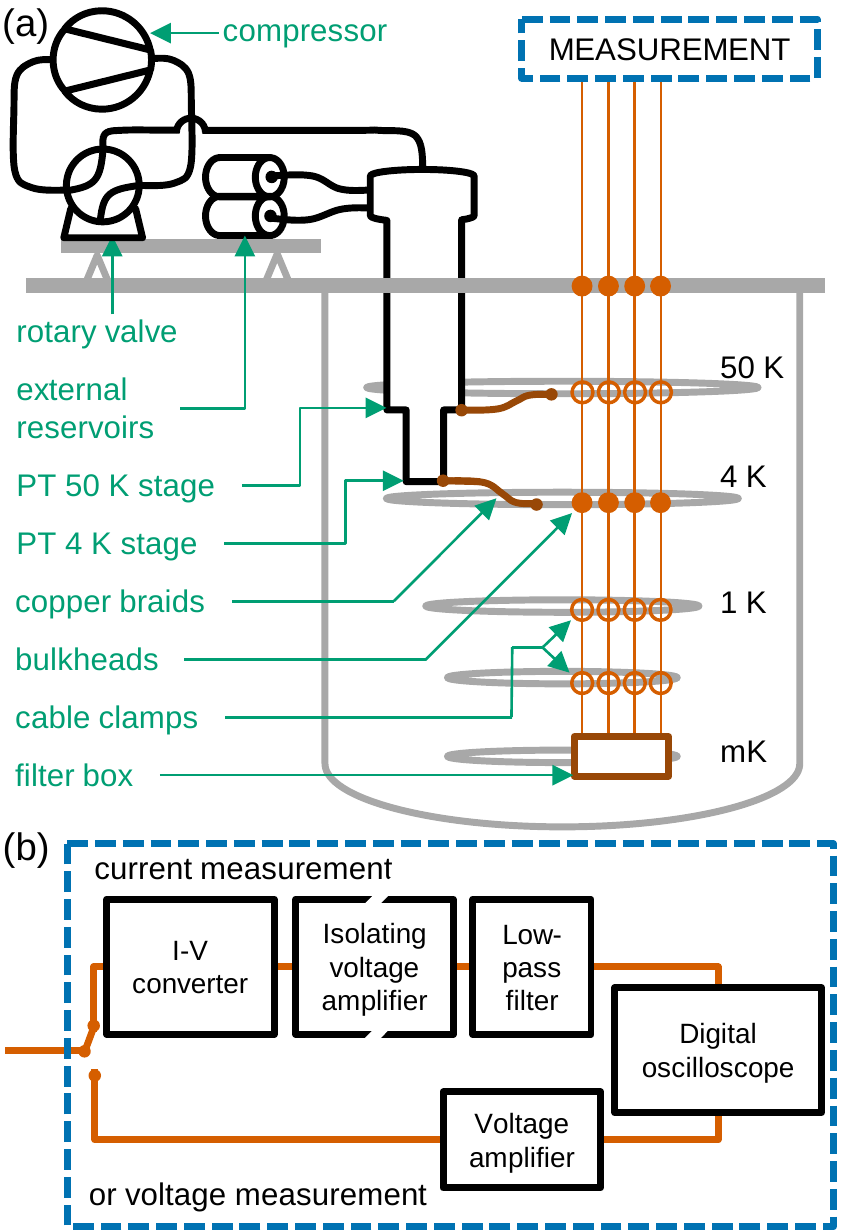}%
	\caption{(Color online) (a) Schematic of the cryogen-free dilution refrigerator measurement setup. The two stages of the pulse tube cooler (black) are connected to the 50~K and 4~K plates inside the cryostat (gray) with copper braids for vibration isolation. The coaxial cables (orange) run from the measurement setup into the cryostat and down to a filter box (brown), thermalized at each flange with either a bulkhead (filled circle) or a clamp (empty circle). (b) Block diagram of the measurement instrumentation used for either current or voltage measurements. In both cases, a digital oscilloscope acquires the final signal.}
	\label{SetupDrawingFig}%
\end{figure}

Figure~\ref{SetupDrawingFig} shows a schematic of our BlueFors BF-LD400 dilution refrigerator setup. Pre-cooling to $<4$~K is provided by a two-stage Cryomech PT-415 pulse tube cryocooler, which is driven by a CP1000 helium compressor. As shown in the schematic, the low and high pressure lines in and out of the compressor are connected to a rotary valve. This has an internal disc that is incrementally rotated at a frequency of 140~Hz by a stepper motor with a full rotation frequency of 1.4~Hz. With each half rotation, the rotary valve switches the connection of the pulse tube between the low and high pressure lines. The valve is mounted on a metal plate, along with the helium gas reservoirs, that stands on rubber posts to isolate the 140~Hz vibrations from the cryostat. The pulse tube is designed to be vibrationally isolated from the cryostat, where soft copper braids are used to thermally link the cold heads to their respective plates in the dilution refrigerator.

The cables run from the `measurement box,' as shown in Figure~\ref{SetupDrawingFig}, from room temperature down to the mixing chamber plate. Our standard setup uses cupronickel semi-rigid cables from COAX CO, part number SC-219/50-CN-CN (we refer to this cable as `UT85 cupronickel'). We have SMA bulkheads from Huber-Suhner at two stages. The hermetically sealed 34\_SMA-50-0-3/111 is used at the top plate and the standard 34\_SMA-50-0-1/111 is used at the 4~K plate. On the 50~K, 1~K and cold plates, we use copper clamps made in-house for thermalization of the cables without the need for bulkheads. Finally, these cables go to a filter box made in-house at the mixing chamber plate. The filter box houses a printed circuit board where, for each line, a Mini-Circuits LFCN-80+ LC-pi low-pass filter cuts off frequencies greater than 80~MHz. This provides effective attenuation up to $\sim 3$~GHz. To filter out noise beyond this frequency, each line is counter-wound around an ECCOSORB rod and the box is filled with copper powder. Not shown in this figure is the space in which the sample is mounted below the mixing chamber plate and the superconducting vector magnet from American Magnetics. Also not shown is a constantan twisted-pair `loom' wire used to carry our low-frequency signals to the device that goes through a separate filter box.

We perform measurements on gated silicon nanostructures for spin-qubit experiments \cite{Morello2010, Pla2012, Pla2013, Muhonen2014}. For these experiments, it is important to minimize vibrations of the sample with respect to the field applied by the superconducting magnet. Furthermore, vibrations can affect spin-qubit measurements through the creation of voltage noise on the gate electrodes of the device and the addition of noise to the measurement signal. The measured signal is the current flowing through a single-electron transistor (SET), which is switched off and on by single-electron tunneling events to and from a nearby $^{31}$P donor. These events occur with a frequency of 100~Hz to 100~kHz and modulate the current with a typical contrast of 1~nA.

Figure~\ref{SetupDrawingFig}(b) shows our typical signal measurement setup. For current measurement, we use the FEMTO DLPCA-200 transimpedance amplifier set to low-noise mode with a gain of $10^7$~V/A, which has an input impedance of 150~$\Omega$ and a passband from DC to 50~kHz. The signal is then amplified by a Stanford Research Systems (SRS) SIM910 voltage amplifier module on a SIM900 mainframe with gain 10~V/V. The input shield is set to floating to break a ground loop. The output is passed through a SIM965 analog filter module set to a low-pass fourth order Bessel filter with a cut-off frequency of 40~kHz and negligible output impedance. An AlazarTech ATS9440 PCI digitizer card acquires the final signal. As shown in the figure, a different instrument is used to perform voltage measurements. While we do not typically measure voltages within the scope of our experiments, this was set up to measure the noise on our lines, as we will discuss later in the paper. For voltage measurements, we use the SR560 voltage amplifier set to a gain of either $10^3$ or $10^4$~V/V, with a low-pass filter at 100~kHz with 12~dB/oct roll-off. The amplifier has an input impedance of 100~M$\Omega$ and output impedance of 50~$\Omega$. The digitizer used to acquire the final signal has an input impedance of 50 $\Omega$ as well, which halves the gain of our voltage amplifier chain. The use of the SR560 here introduces a ground loop in the system, but we focus on the spectral components of interest.

\section{Pulse tube noise}
\label{sec:PTnoise}

\begin{figure}[t]
	\includegraphics[width=\columnwidth, keepaspectratio = true]{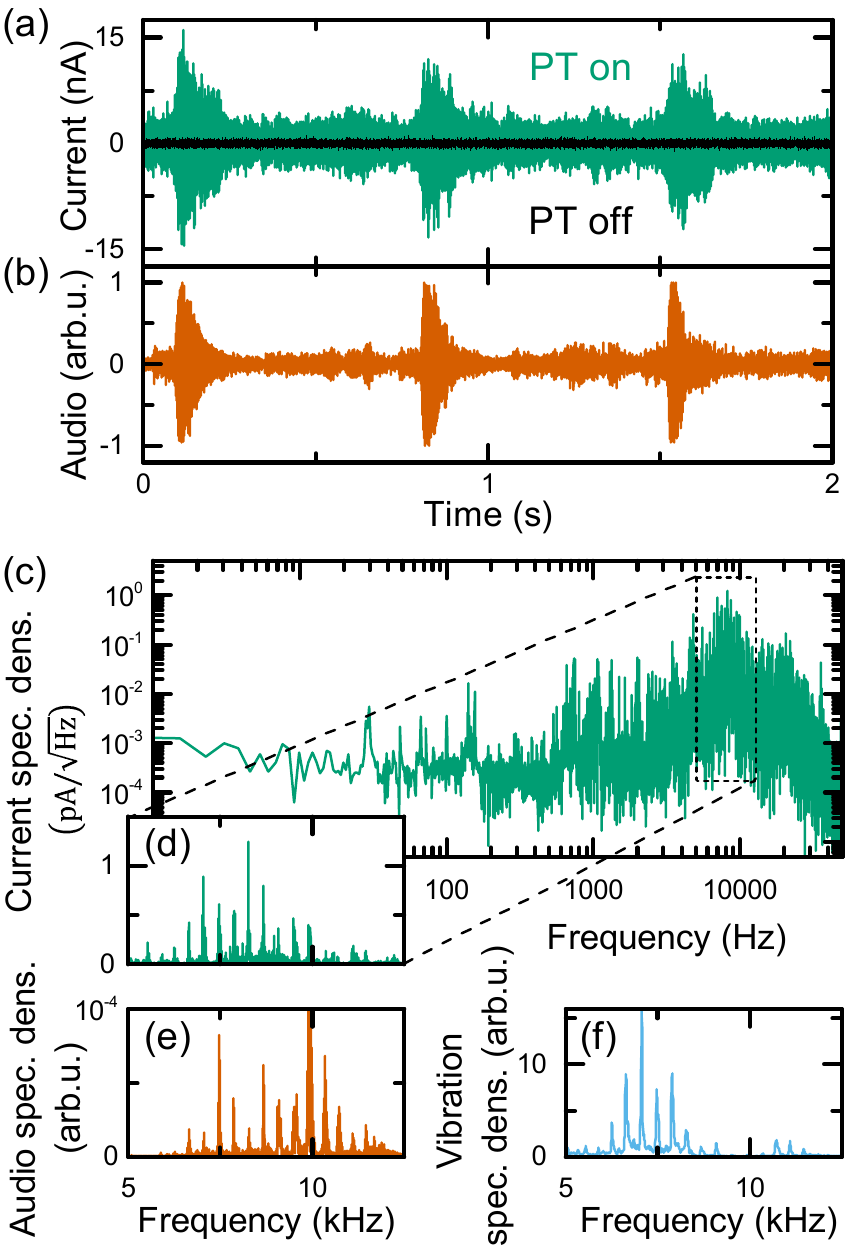}%
	\caption{(Color online) (a) Current traces measured on a UT85 cupronickel cable run from room temperature to the sample at 10 mK. The green and black traces are taken when the pulse tube is on and off, respectively. (b) Audio recording of the pulse tube chirps. (c) Amplitude spectral density of the current trace with a zoom-in of the 5-12~kHz range shown in (d). (e) Amplitude spectral density of the audio trace. (f) Vibration spectral density measurement taken with an accelerometer.}%
	\label{NoiseIntroFig}
\end{figure}

With the system at its base temperature of $\sim11$~mK, we set up our measurement of the current through the SET. A 2-second trace of the current acquired with 200~kSa/s is shown in Figure~\ref{NoiseIntroFig}(a), labeled `PT on.' The gain of the amplifier chain ($0.1$~V/nA) has been taken into account to plot the magnitude of the current. We observe noise that coincides with the chirping of the pulse tube. Chirping refers to the audible 1.4~Hz cycle of connecting the pulse tube to the high and low pressure lines from the compressor. With an amplitude exceeding 10 nA, the noise would completely overwhelm our experimental signal. We temporarily switch off the pulse tube cooler, such that both the compressor and the rotary valve are switched off, without interrupting the dilution refrigeration to acquire the trace labeled `PT off.' The noise amplitude reduces by a factor of 8, clearly indicating that the noise is caused by the operation of the pulse tube.

Figure~\ref{NoiseIntroFig}(b) shows an audio recording of the chirping next to the setup, taken with the standard sampling frequency of 44.1~kHz. Note that traces (a) and (b) were not taken simultaneously. Figure~\ref{NoiseIntroFig}(c) shows the amplitude spectral density of the current trace. The dominant noise spectrum is between 5 and 10~kHz, which overlaps with the bandwidth of the spin readout experiment we typically perform \cite{Morello2010}. We also observe a peak at 140 Hz, caused by vibrations of the rotary valve, but this is negligible compared to the dominant noise.

Figure~\ref{NoiseIntroFig}(d) shows a zoom-in of the dominant noise. We observe sharp peaks in the 5 to 10~kHz range that are regularly spaced with a separation of $\sim400$~Hz. We also calculate the amplitude spectral density of the audio signal and show a zoom-in of the same frequency region in Figure~\ref{NoiseIntroFig}(e). We observe the same pattern, with the frequency peaks matching those from the current measurements. In another test, we took vibration measurements with a Wilcoxon Research 731A seismic accelerometer. While this model is low-pass filtered at 450~Hz, we still managed to acquire an appreciable signal in the kHz range. We placed the accelerometer at the top of the frame on which the fridge is mounted. Figure~\ref{NoiseIntroFig}(f) shows the amplitude spectral density of the vibration velocity in the same frequency range for an average of 100 traces. We observe the same characteristic 400~Hz spacing between peaks, again with the frequencies matching the previous traces.

\begin{figure*}
	\includegraphics[width=\textwidth, keepaspectratio = true]{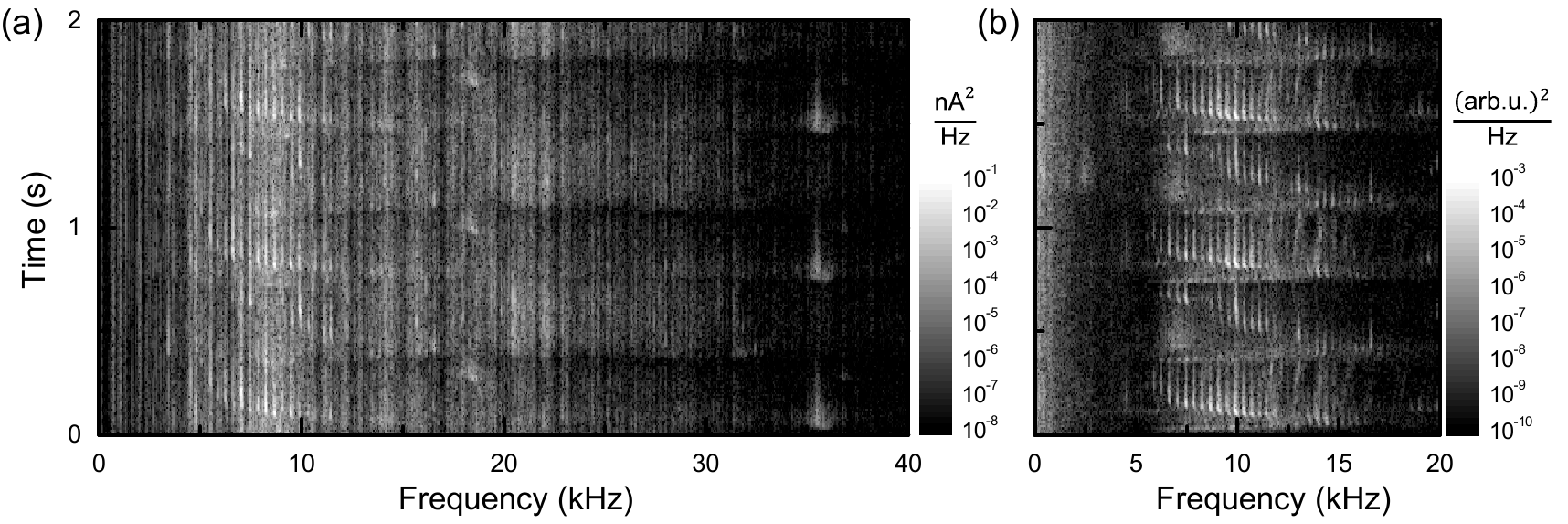}%
	\caption{Spectrograms of (a) the current trace shown in Figure~\ref{NoiseIntroFig}(a), and (b) the audio record shown in Figure~\ref{NoiseIntroFig}(b).}
	\label{SpectroFig}%
\end{figure*}

Figure~\ref{SpectroFig} plots the spectrograms of both the current and audio signals to reveal how their spectra vary with time. The spectrograms are obtained by dividing the signals into 20~ms windows, with an overlap between adjacent windows of 10~ms for smoothing. A Hamming window is used to reduce spectral leakage. The power spectral density is then calculated from the Fourier transform, and the spectrograms are plotted to 40~kHz for the current signal in Figure~\ref{SpectroFig}(a) and to 20~kHz for the audio signal in Figure~\ref{SpectroFig}(b). The spectrograms reveal how the regularly spaced peaks in the 5 to 10~kHz range from Figure~\ref{NoiseIntroFig}(d)-(f) are produced. Figure~\ref{SpectroFig} shows three chirps of the pulse tube, occurring at times $\sim 0.1$~s, $\sim 0.8$~s and $\sim 1.5$~s. We see that with each chirp, the 400~Hz-spaced peaks are not produced simultaneously, but rather each peak occurs in quick succession. Each subsequent peak is spaced by roughly 400~Hz in frequency and by 10 to 20~ms in time. Following a peak, there appears to be a finite decay time in the magnitude of noise at that frequency. This is shown more clearly in the spectrogram of the audio signal, where the result is a `slanted comb' structure. This comb also appears in between chirps, at $\sim 0.5$~s and $\sim 1.2$~s, presumably when the rotary valve connects the low pressure line from the compressor. It is interesting that these combs are slanted in the same direction as the ones where the high pressure line is connected, although they have a lower amplitude and a longer delay between successive peaks.

With the understanding that the observed electrical noise originates from (or is at least strongly correlated with) vibrational noise, we perform further tests to gain insight into the problem. We first test the effect of ambient acoustics by playing a 3~kHz tone next to the cryostat to see if it couples to the measured signal. With the volume set much louder than the chirps of the pulse tube, the amplitude spectrum still shows that the magnitude of the tone is significantly lower than the 5 to 10~kHz peaks (data not shown). This suggests that the problematic vibrations are not coupling via acoustics through the air to the top flange. We then attempt to reduce the noise by replacing the default step driver to the rotary valve with a linear ramp driver. We try this with the LNII linear micro-stepping drive from Precision Motion Controls. While this slightly reduces the amplitude of the 140~Hz peak as measured both by the accelerometer and the current signal, it does not make a difference to the dominant noise in the 5 to 10~kHz range (data not shown).

Next we test the hypothesis that the vibrations may be originating from the gas reservoir and rotary valve, combined with ineffective damping from the rubber posts supporting that plate. We therefore detach the entire plate and lift it a few cm over the frame. The helium lines are, of course, still connected to the pulse tube. Analyzing the amplitude spectra of the two traces, this does not make any difference to the dominant 5 to 10~kHz noise observed. However, the 140~Hz peak caused by the vibrations of the rotary valve reduces in amplitude by a factor of 11 (data not shown). This led us to investigate whether the 140~Hz peak actually couples in to the room-temperature part of the measurement setup as opposed to the cable in the cryostat. Careful positioning of the room-temperature connection to the transimpedance amplifier reduces the 140 Hz peak close to the noise floor, while the higher frequency noise associated with the pulse tube chirps remains unaffected. This supports the hypothesis that the 140 Hz peak is being coupled through the room-temperature part of the setup.

\begin{figure}[t!]
	\includegraphics[width=\columnwidth, keepaspectratio = true]{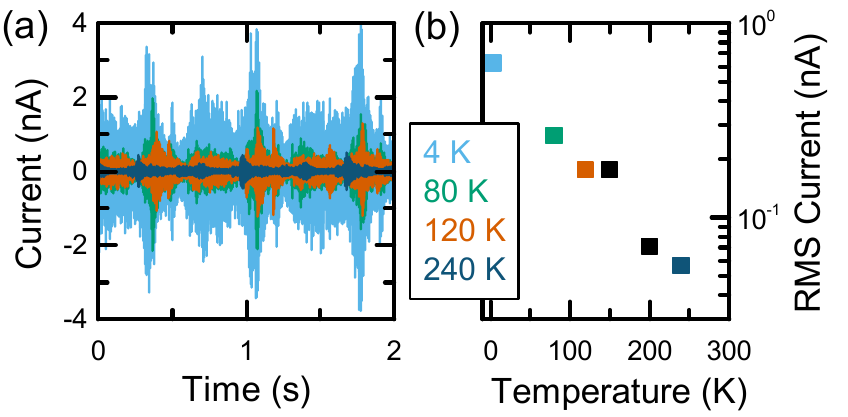}%
	\caption{(Color online) The current is measured at different temperatures of the second stage of the pulse tube during a cool-down. The UT85 cupronickel cable is run from the room-temperature flange to the 4~K plate. (a) The current traces are plotted for four selected temperatures. (b) The RMS current is plotted as a function of temperature.}
	\label{TemperatureFig}%
\end{figure}

Finally, we investigate the dependence of the magnitude of the noise on the temperature of the system. For this test, we use a UT85 cupronickel line that terminates at an open-circuit bulkhead at the 4~K plate. Figure~\ref{TemperatureFig}(a) plots the current traces at different temperatures, clearly showing the increase in the magnitude of the noise with decreasing temperature. Figure~\ref{TemperatureFig}(b) plots the root-mean-square (RMS) values of the current traces as a function of temperature, suggesting an exponential dependence. 

While the pulse tube chirps reduce by over an order-of-magnitude from 4~K to 240~K, the 140~Hz peak reduces by less than a factor of two. This also supports the hypothesis that the 140 Hz vibrations are largely coupling in to the signal via the room-temperature cables.

The dependence of the magnitude of the pulse tube chirps on temperature must come from the origin of the vibrations within the pulse tube itself and/or the mechanism by which they couple to the electrical signal. The volume of the sound made by the pulse tube during the cool-down does not vary significantly. If this is a good measure of the magnitude of the vibrations at the source, then this would suggest that the temperature dependence lies in the coupling mechanism.

The amplitude of the current noise of the trace shown in Figure~\ref{NoiseIntroFig}(a), where a cable reaching the device enclosure was tested, is approximately 3 times larger than that of the 4~K trace shown in Figure~\ref{TemperatureFig}(a), where the cable stops at the 4~K plate. The dependence of the noise magnitude with cold cable length is consistent with vibration-to-signal coupling occurring within the cables. We have two models for how vibrations can couple to the electrical signal, both of which are based on the fact that the dielectric undergoes a greater thermal contraction than the outer conductor.

The dielectric should undergo the majority of its thermal contraction at a relatively high temperature. From Figure~\ref{TemperatureFig}(b), however, the noise appears to continually increase with decreasing temperature down to 4~K.  Recall that the cable tested is fixed on one end to the room-temperature plate and on the other to the 4~K plate, resulting in a thermal gradient. Therefore, the length of the cable that is cold, and that can therefore translate vibrations to electrical noise, continually increases with decreasing temperature at the cold end.

The first candidate model is the creation of noise due to microphonics. We refer specifically to the effect used in a capacitor microphone, where acoustic vibrations cause fluctuations in capacitance. A current is produced that is proportional to these fluctuations and the voltage applied across the capacitor. In our cables, the thermal contraction of the dielectric causes it to be squeezed around the inner conductor and results in a gap being formed between it and the outer conductor. Thus, vibrations can cause movement of the dielectric and inner conductor with respect to the outer conductor, which translates to variations in the effective capacitance of the line. While we have not intentionally applied a voltage on the line, one may be created by the thermoelectric effect. A voltage difference on the order of millivolts would be induced between the cold and room temperature ends of the line, and a difference may be present between the inner and outer conductors depending on their construction and assembly.

We test if the dominant mechanism is microphonics by applying a voltage on the line while measuring the current, which can be achieved with the FEMTO DLPCA-200 transimpedance amplifier. Applying biases of up to 5~V on the line, we find that the current measured is independent of the applied bias (data not shown). Based on this, we believe that the dominant effect is not capacitor microphonics.

The second candidate is the creation of noise via the triboelectric effect \cite{Ong1987}. When the dielectric is rubbed against either the inner or outer conductor, any friction experienced will facilitate the transfer of charge that can be measured as either a current or a voltage depending on the setup. The shrinking of the dielectric with decreasing temperature allows more room for sliding and rubbing between the dielectric and outer conductor. The amount of friction and charge transfer inside a cable will vary significantly with the construction of and materials in the cable. Therefore, we proceed by installing and measuring several different types of cables in our setup, which will be discussed in detail in the following section.

\section{Cable Comparison}
\label{sec:CableComparison}

We prepare a measurement of the noise on different types of cables with the aim of understanding the coupling mechanism and of identifying ways to reduce the noise as much as possible. We test four variations of semi-rigid cables and four variations of flexible cables. For fair comparison, all of these cables run from room temperature to open-circuit terminated bulkheads at the 4~K plate. The semi-rigid cables tested are clamped at the 50~K stage for thermalization as described before. The flexible cables are thermalized by taping and tying them to posts both above the 50~K and 4~K plates. The system is cooled to $<4$~K and the dilution unit is not operated. We record both the voltage and current noise present on the cables. Measurements are performed as described previously.

\begin{figure*}
	\centering
	\includegraphics[width=\textwidth]{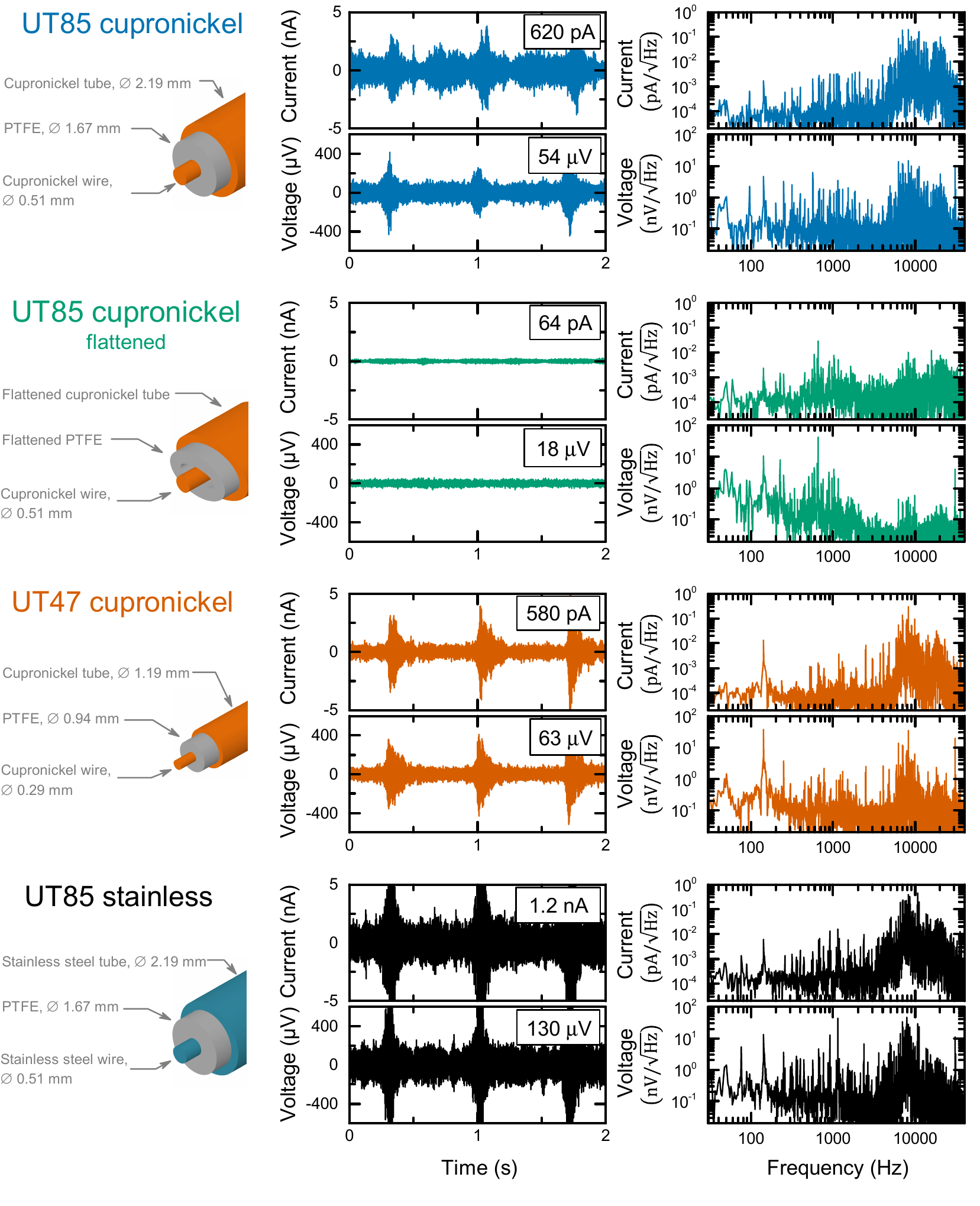}
	\caption{(Color online) Each row shows the noise measurements for a particular semi-rigid cable, terminating in an open-circuit at a bulkhead on the 4~K plate. A schematic of the cable is shown in the first column. The second column shows the current and voltage traces, with their RMS values written in the top-right corner of each graph. The third column shows the amplitude spectral density of the current and voltage traces.}%
	\label{CompareCableFigSR}
\end{figure*}

\begin{figure*}
	\centering
	\includegraphics[width=\textwidth]{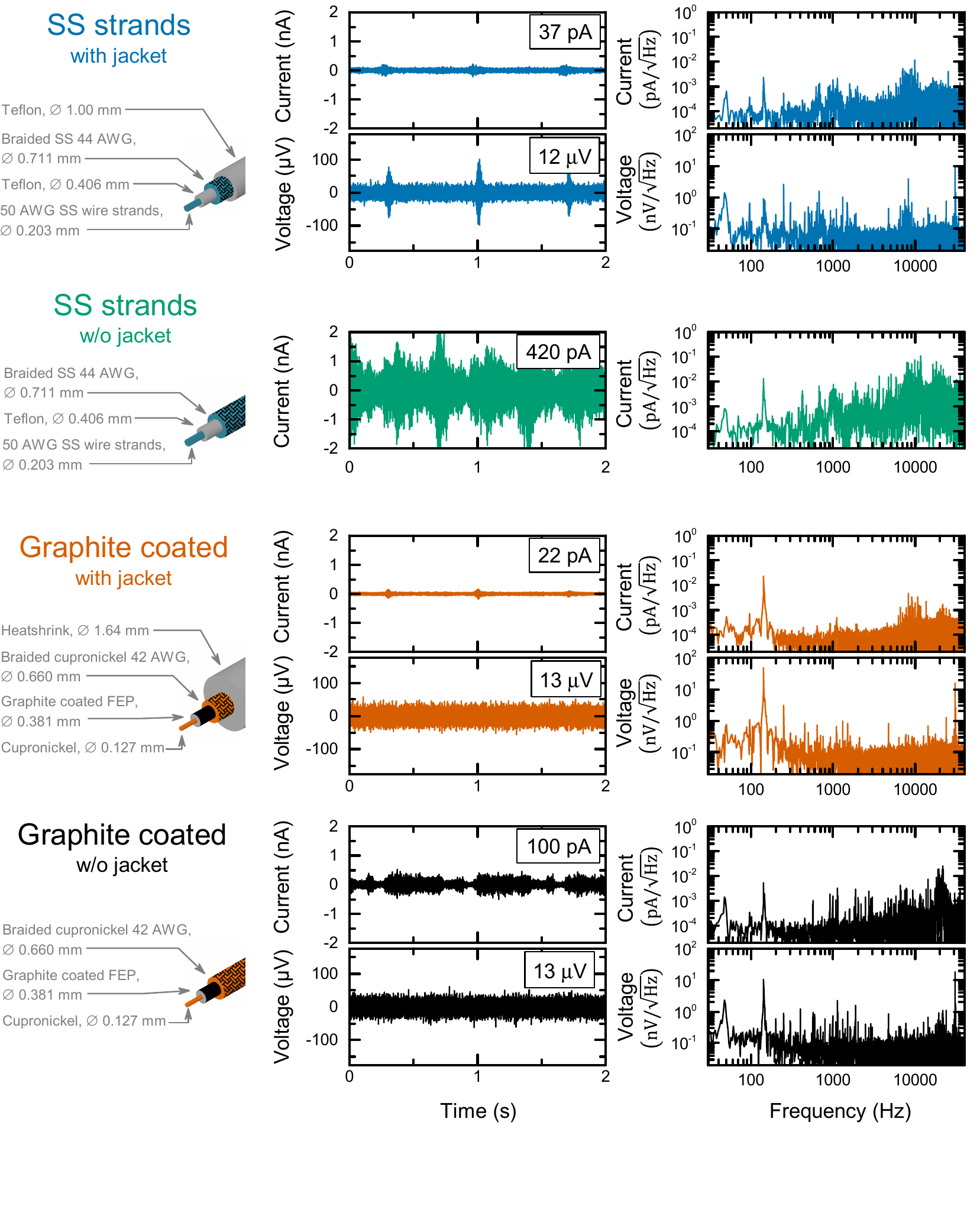}
	\caption{(Color online) Each row shows the noise measurements for a particular flexible cable, terminating in an open-circuit at a bulkhead on the 4~K plate. A schematic of the cable is shown in the first column. The second column shows the current and voltage traces, where available, with their RMS values written in the top-right corner of each graph. The third column shows the amplitude spectral density of the current and voltage traces.}%
	\label{CompareCableFigFlex}
\end{figure*}

Figures~\ref{CompareCableFigSR} and~\ref{CompareCableFigFlex} show the results for the semi-rigid and flexible cables tested, respectively. The first column shows a schematic of the cable, with dimensions and materials labeled. The second column shows the time traces of the current and voltage noise, where available. The traces plotted here have been filtered digitally to remove components that are not representative of the noise created in the cable. A notch filter is used to remove the 140~Hz signal that couples in to both the current and voltage measurements in the room-temperature setup. Additionally, the voltage measurement introduces a peak at $\sim$30~kHz independent of the cable, which we also remove with a notch filter. The RMS values labeled in the figure are calculated using the filtered traces. The third column shows the amplitude spectral density of the unfiltered current and voltage traces, so that the 140~Hz and 30~kHz peaks are visible here. Note that the voltage and current noise measurements are not normalized to the length of the cable, as they are installed with the necessary length to reach the 4~K plate with the thermalization steps in our setup.

\begin{figure}[t!]
	\includegraphics[width=\columnwidth, keepaspectratio = true]{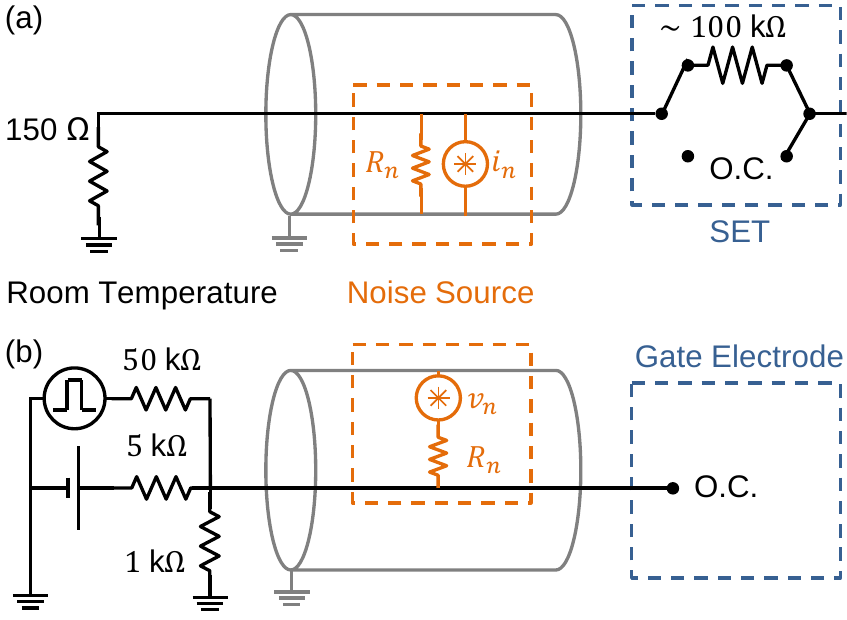}%
	\caption{(Color online) (a) Schematic of the current measurement setup with the Norton equivalent of the noise source shown. (b) Schematic of the setup for applying DC and AC voltages to a gate electrode with the Thevenin equivalent of the noise source shown.}
	\label{NoiseAndSetupFig}%
\end{figure}

The first cable presented in Figure~\ref{CompareCableFigSR} is the original cable tested, the semi-rigid UT85 cupronickel cable from COAX CO. The RMS values for the current and voltage noise traces are 620~pA and 54~$\mu$V. From this, we estimate the effective output impedance, $R_n$, of this `noise source' between the inner and outer conductors of the cable to be 90~k$\Omega$. This is an important quantity in determining whether or not this noise is detrimental to a particular experiment. We thus analyse our setup in relation to the measurement of the current through the SET and on the voltage biases applied to electrodes on the device. The setup for measuring the SET current is as described before and is summarized in Figure \ref{NoiseAndSetupFig}(a). A transimpedance amplifier with an input impedance of 150~$\Omega$ sits at room temperature and is connected to a cable that runs inside the cryostat. The output of the cable is connected to an LC-pi 80~MHz low-pass filter at the mixing chamber plate which can be ignored in the context of the $\sim$kHz pulse tube noise, and is thus omitted from Figure \ref{NoiseAndSetupFig}. The output of the filter is then connected to the drain contact of the SET. The impedance of the SET to the source contact varies between $\sim$100~k$\Omega$ and an open-circuit, depending on whether it is biased to a conductance peak or in Coulomb blockade. This means that the noise source at the cable `sees' a high impedance to its right (the SET) and a low impedance to its left (the transimpedance amplifier). Therefore, the majority of the noise current $i_n$ flows through the transimpedance amplifier, corrupting the experimental signal of the current through the SET. Techniques such as lock-in modulation and RF reflectometry \cite{Angus2008} are able to modulate the readout signal to a higher frequency, thus avoiding the $\sim 10$ kHz pulse tube noise. However, care must still be taken in considering the effect of this noise on the voltage biasing of the gate electrodes.

The experiments we conduct are sensitive to voltage noise present at the gate electrodes of the device \cite{Laucht2015}. The gates we intend to pulse with moderately high frequencies are connected through 80~MHz low-pass filters which, again, can be ignored in the context of the $\sim$kHz pulse tube noise. At room temperature, a passive voltage summer-divider circuit, with resistors in the order $\sim$k$\Omega$, is used to combine a DC voltage from an SRS SIM928 isolated voltage source and voltage pulses from a Teledyne-Lecroy ArbStudio 1104 arbitrary waveform generator. This means that the noise source at the cable `sees' an open circuit to its right (the gate electrode) and a resistance of a $\sim$k$\Omega$ to its left (the output of the room temperature summer-divider circuit). In this configuration, the voltage noise at the gate electrode would only be a few percent of the voltage noise $v_n$ produced by the cable, since $R_n$ is much larger than the output impedance of the voltage divider-combiner.  In Section~\ref{sec:coherence}, we quantify the effect of this noise on our experiments.

The second row of Figure~\ref{CompareCableFigSR} shows the noise measured on an identical cable that has been flattened with a vise. The entire length of the cable is flattened apart from the sections that are looped for strain relief and thermalization of the inner conductor. The purpose of the flattening is to restrict movement of the dielectric when the cable is cooled to reduce triboelectric effects. Comparing the RMS values for the current and voltage traces, the flattening yields an improvement of factor 10 and 3, respectively. From the amplitude spectra, we see that the pulse tube noise near 10~kHz has reduced by an order of magnitude for both current and voltage measurements. While the current signal for the standard cable is dominated by the spectral components in this region, the voltage noise has lower frequency contributions with comparable amplitude. The noise in the lower frequency region (30 to 3000~Hz) is not reduced by flattening the cable. 

The third and fourth rows of Figure~\ref{CompareCableFigSR} show the noise measured on a thinner cable made of the same material (UT47 cupronickel) and a cable of the same dimensions made with stainless steel (UT85 stainless), both from COAX CO. The thinner cable is the SC-119/50-CN-CN and the stainless steel cable is the SC-219/50-SS-SS. The noise amplitudes are of the same order as that of the standard UT85 cupronickel cable.

Figure~\ref{CompareCableFigFlex} shows the noise measurements on four types of flexible cables. The first row shows the noise measured on the `ultra miniature coaxial cable' Type SS from Lake Shore Cryotronics. This cable, as the drawing shows, has stainless steel strands for the center conductor, Teflon insulation, braided stainless steel wire as the outer conductor, and a Teflon jacket. The current and voltage noise is significantly lower as compared to the standard semi-rigid cables, with RMS values of 37~pA and 12~$\mu$V, respectively. We conjecture that the Teflon jacket serves to squeeze the braided outer conductor onto the dielectric to prevent movement within the cable when cooled. We thus take an identical cable and strip it of the jacket, as shown in the second row of the figure. The current trace has an RMS value of 420~pA, which is a factor of 11 worse than the jacketed version. The amplitude spectrum shows that this order of magnitude increase in the noise is across the entire frequency range. The record for the voltage measurement is not available for this cable. 

The final cable we test is a graphite-coated cable given to us by David Goldhaber-Gordon. In their paper on vibrational noise in a pulse-tube system, Pelliccione \textit{et al.} reported the use of a cable with graphite coating on the outside of the dielectric to reduce triboelectric effects due to rubbing against the outer conductor \cite{Pellicione2013}. The graphite is a conductive solid-state lubricant that is meant to reduce friction between the dielectric and outer conductor and to rapidly return any charges displaced due to triboelectrics to and from the outer conductor \cite{Ong1987}. As stated in their paper, this cable from Calmont Wire \& Cable comprises of a single 36~AWG cupronickel wire for the inner conductor and braided cupronickel for the outer conductor. The dielectric is FEP with graphite coating on the outside. This cable did not come with an outer jacket. To further validate our hypothesis that the jacket helps to reduce triboelectric effects, we made a jacket for the cable using $1.5/0.5$~mm heat-shrink tubing from HellermannTyton. A schematic of the resulting cable is shown in the third column of Figure~\ref{CompareCableFigFlex}. The current noise measured is the best so far. The RMS current is 22~pA and the peak amplitude during the chirps of the pulse tube is less than 200~pA. Note that the heat-shrink that was available to us becomes brittle at low temperatures. Although we do not know the degree to which it thermally contracts before becoming brittle, a Teflon jacket made by the cable manufacturer should improve the noise performance of the cable further.

We then test an identical cable without the heat-shrink jacket, shown in the last column of Figure~\ref{CompareCableFigFlex}. The magnitude of the noise is significantly increased, as expected, with an RMS current of 100~pA. However, the time trace looks quite different compared to the other cables tested. The amplitude spectrum reveals that the dominant noise is actually close to 20~kHz, as opposed to the usual peaks in the 5 to 10~kHz range. Surprisingly, the amplitudes of the peaks in the 5 to 10~kHz range for the jacketed and unjacketed graphite-coated cables are similar. A spectrogram of the current trace of the unjacketed cable (data not shown) reveals that the noise around 20~kHz does not appear in peaks, but as a smear, similar to the spectrogram shown in Figure~\ref{SpectroFig}(a) for the semi-rigid UT85 cupronickel cable. The same pattern is observed in comparing the voltage spectra of the jacketed and unjacketed cables. While the voltage traces look similar and both have RMS values of 13~$\mu$V, the noise near 20~kHz can be seen appearing above the noise floor for the unjacketed cable. We do not understand why the jacketing of the graphite cable reduces this 20~kHz noise, but does not affect the 5 to 10~kHz peaks.

\begin{figure}
	\centering
	\includegraphics[width=\columnwidth]{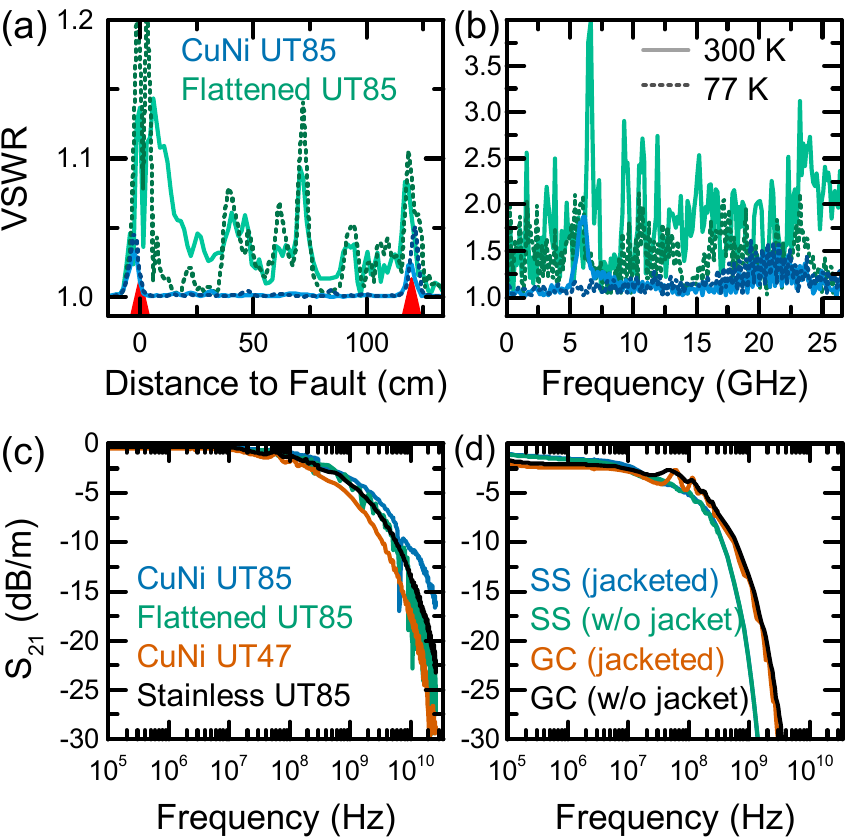}
	\caption{(Color online) Panels (a) and (b) show the voltage standing wave ratio (VSWR) for the UT85 cupronickel standard (blue) and flattened (green) cables (a) as a function of distance to fault (DTF) and (b) as a function of frequency. The solid lines refer to measurements recorded at 300 K, and the dotted lines to measurements recorded at 77 K. The red markers in (a) indicate the start and the end of the tested cable. The frequency response (S$_{21}$) is plotted for all the (c) semi-rigid and (d) flexible cables tested in this paper. The measurements are performed at room-temperature and the results are normalized to the length of the cable. `GC' in (d) refers to the graphite-coated cable. }%
	\label{S21Fig}
\end{figure}

For the reader's reference, Figure~\ref{S21Fig}(a-b) plots the voltage standing wave ratio (VSWR) of the standard and flattened UT85 cupronickel cables at room temperature (solid lines) and at 77 K (dotted lines). These measurements are taken with the Keysight Technologies N9918A FieldFox hand-held microwave analyzer. The results are plotted as a function of the distance to fault in sub-figure~(a) and as a function of frequency in sub-figure~(b). As expected, the VSWR is generally close to unity for the standard cable. The VSWR for the flattened cable is between 1 and 4 at room temperature and between 1 and 2 at 77 K. Note that the cable was flattened with a vise in a piecewise manner, whereas a uniformly flattened cable should perform much better. Also for the reader's reference, Figure~\ref{S21Fig}(c-d) plots the frequency response (S$_{21}$) at room-temperature of all the cables tested in this paper. The S$_{21}$ results for the semi-rigid cables and flexible cables are plotted in sub-figure~(c) and~(d), respectively, normalized to the length of the cable. From sub-figure~(c), we see that the flattening of the UT85 cupronickel cable does not significantly worsen its S$_{21}$ frequency response. The roll-off above 1~GHz is slightly steeper and there are some ripples in the passband beyond 100~MHz. From Figure~\ref{S21Fig}(d), we see that the frequency responses of the jacketed and unjacketed stainless steel cables are identical. This is expected, as the unjacketed cable was obtained simply by stripping off the jacket. However, this is not the case for the graphite-coated cables. While the overall passband and roll-off are the same, the jacketed cable has some ripples beyond 30~MHz. This could be due to deformation of the dielectric during the heating of the cable to secure the heat-shrink jacket. A custom cable with a proper Teflon jacket inserted by the manufacturer should avoid this problem.

\section{Effect of noise on qubit coherence}
\label{sec:coherence}

Following the insights obtained from our cable performance comparison, we prepare the cryogen-free dilution refrigerator for measurement of a well-characterized spin qubit device. The purpose of this is to determine whether or not the new setup allows for successful measurement of the device, given our efforts in mitigating pulse tube noise. We then perform coherence measurements to measure the effect of this noise on the electron spin qubit.

We use the device labeled `Device B' in Reference~\onlinecite{Muhonen2014}, which was previously measured in an Oxford Kelvinox 100 wet dilution refrigerator, in order to perform a direct comparison between the two different dilution refrigerator setups. The qubit is the electron spin of a single $^{31}$P donor in isotopically purified $^{28}$Si.

Operation of this device requires four different types of lines: (1) low-frequency lines, which maintain a constant voltage over the period of an experiment; (2) pulsing lines, which need to be pulsed at frequencies on the order of $\sim$1~MHz; (3) source-drain lines, which are used to detect current pulses on the order of $\sim$100~kHz; and (4) a broadband line, used to send 10~MHz--50~GHz excitations to perform spin resonance on the phosphorus atom. The low-frequency lines consist of constantan twisted-pair `loom' wire running from room temperature to 4~K, where they go through a filter box similar to the one described in Section~\ref{sec:Setup}, but with a 100~Hz second-order RC low pass filter replacing the 80~MHz filter. For the pulsing lines and source-drain lines, we use flattened semi-rigid UT85 cupronickel cables and jacketed graphite-coated cables, respectively. We do not use graphite cables for the pulsing lines due to lack of cable at the time of the experiment. The broadband line consists of a UT85 silver plated cupronickel coaxial cable (219/50-SCN-CN from COAX CO) running from room temperature to the device enclosure. Thermalization of this line is achieved via a 10~dB attenuator at the 4~K plate and a 3~dB attenuator at the mixing chamber plate. The line also contains a double DC block at room temperature and an inner-only DC block at the mixing chamber, both with nominal cut-off frequencies of 10~MHz. While this cable will exhibit a significant degree of triboelectric noise (Section~\ref{sec:PTnoise}), we currently do not have an alternative solution capable of delivering the $\sim40$ GHz signals required for qubit control.

As expected, the modified setup with low-noise cables allows us to successfully tune and measure the device in the BlueFors BF-LD400 dry fridge, without noticeably affecting the measurement fidelity of the qubit. We then proceed to qubit coherence measurements, which allow probing of the environmental noise that causes dephasing of the electron spin. By measuring its coherence times ($T_2$) using carefully designed dynamical decoupling sequences, we can extract the power spectral density of the noise affecting the qubit, across the frequency range relevant to the qubit operation and coherence times \cite{Alvarez2011}. This noise spectrum has been previously measured in this device inside the Oxford Kelvinox 100 wet dilution refrigerator \cite{Muhonen2014}. We choose the parameters of the noise spectroscopy pulse sequences to map the noise spectrum over the range 1 to 25~kHz, which coincides with the bandwidth over which we have been studying the pulse tube noise.

\begin{figure}
	\centering
	\includegraphics[width=\columnwidth]{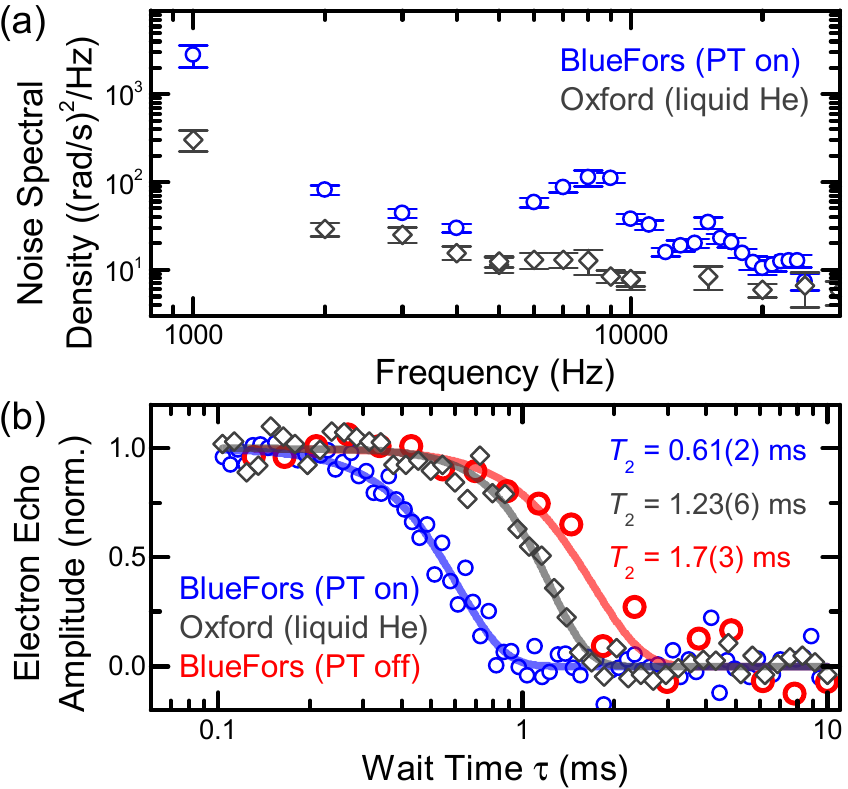}
	\caption{(Color online) Coherence measurements of an electron spin qubit in different dilution refrigerator setups. In both panels, diamonds correspond to measurements taken inside an Oxford Kelvinox 100 liquid He bath dilution refrigerator, while circles correspond to measurements taken in a BlueFors LD400 cryogen-free dilution refrigerator. (a) Noise spectral density of the qubit, extracted from dynamical decoupling coherence time measurements. (b) Hahn echo coherence time measurements. The two dry fridge measurements correspond to normal operation (blue) and dilution unit operating while the pulse tube is switched off (red). The time over which the pulse tube was switched off to obtain the latter measurement was $\sim$1~minute.}%
	\label{CoherenceFig}
\end{figure}

Figure~\ref{CoherenceFig}(a) plots the measured noise spectral density as a function of frequency in this setup (blue circles). The figure also reproduces the data from the wet fridge measurements from Reference~\onlinecite{Muhonen2014} for comparison (grey diamonds). We first note an increased level of lower-frequency noise around 1~kHz in the dry fridge setup as compared to the wet fridge setup. This lower-frequency $f^{-\alpha}$-type noise was attributed to movement of the spin inside the inhomogeneous field of the superconducting magnet in Reference~\onlinecite{Muhonen2014}, which suggests that the dry fridge setup suffers from a greater magnitude of these vibrations. Secondly, the dry fridge measurements clearly show the signature of the pulse tube vibrations studied in the first half of this work. We see a greater level of noise in the 5 to 10~kHz region, coinciding with the peaks observed in the characterization presented earlier. As the spectral resolution of our spectroscopy technique is limited, the sharp peaks presented in earlier sections appear here as a broad peak centered at 8~kHz. This is a first indication that the pulse tube vibrations are coupling into the resonance frequency of the electron spin qubit. This is confirmed by a second experiment consisting of a quick Hahn echo $T_2$ measurement \cite{Hahn1950}, taken with the pulse tube temporarily switched off, without interrupting the operation of the dilution unit. Figure~\ref{CoherenceFig}(b) shows the echo decay traces with coherence times of $T_2 = 1.7(3)$~ms and $T_2 = 0.61(2)$~ms for the pulse tube being off and on, respectively. Therefore, we conclude that noise induced by the pulse tube reduces the Hahn echo $T_2$ by a factor $\sim$3. For comparison, the Hahn echo coherence time measured in the Oxford Kelvinox 100 wet dilution refrigerator was $T_2 = 1.23(6)$~ms.

There are three possible mechanisms through which the vibrations caused by the pulse tube can couple to the resonance frequency of the qubit, thus limiting its coherence. The first is based on the fact that vibrations from the pulse tube create large current noise on the coaxial cable used for the broadband line via the triboelectric effect. The current noise created in the section of the cable above the mixing chamber plate will be attenuated by the 3~dB attenuator and the inner-only DC block at that plate before reaching the device. The attenuation of the $\sim10$~kHz noise of the pulse tube achieved by the DC block at base temperature is of order 80~dB. However, another section of cable connects the bottom of the mixing chamber plate to the transmission line on the device, designed to deliver the $\sim40$~GHz signals to the spin qubit. From the perspective of the current noise created in this portion of the cable, the lowest impedance shunting the majority of the current is exactly the short-circuit microwave antenna adjacent to the qubit. Therefore, the vibrations of the pulse tube couple into the magnetic field experienced by the spin qubit via the creation of current noise that then flows through the antenna. Ideally, this would not have an effect on the coherence of the qubit as the magnetic field created would be exactly perpendicular to the quantization axis set by the large externally applied magnetic field. However, due to constraints in device design, a small fraction ($\sim 1/30$) of the amplitude of the field created by the antenna is actually parallel to the quantization axis, and can therefore contribute to dephasing. Using this ratio, we can estimate the noise power experienced by the qubit. Based on Figure~\ref{SpectroFig}, the average power spectral density of the current noise in the 5 to 10~kHz range is of order $10^{-1}$~$\mathrm{nA}^2/\mathrm{Hz}$. The resulting magnetic field noise at the site of the qubit is estimated by approximating the short-circuit antenna as an infinitely-long wire 100~nm away from the donor. Using the electron's gyromagnetic ratio (28~GHz/T), this is converted to noise experienced by the qubit with magnitude $\sim 10$~$\mathrm{(rad/s)}^2/\mathrm{Hz}$. This value is comparable to the observed noise power in the 5 to 10~kHz range (Figure~\ref{CoherenceFig}(a)), indicating that this mechanism may have had a significant contribution. In future experiments, an additional DC block and attenuator should be inserted directly at the device enclosure to filter pulse tube noise created from all sections of the cable. Additionally, the sample should be reoriented to eliminate the component of the field produced by the antenna that is parallel to the quantization axis. 

Another mechanism that could contribute to coupling vibrations to magnetic noise is via the displacement of the device in an inhomogeneous magnetic field. Despite our best efforts, the exact location of the phosphorus atom might not be exactly in the center of the field, causing it to reside in a region where a nonzero magnetic field gradient exists. Mechanical vibrations, of either the magnet or the cold finger to which the device is attached, would thus directly result in magnetic field noise experienced by the qubit. For a qubit located within $\sim$100~$\mu$m of the center of the field, we calculate a maximum field gradient of $\sim5$~mT/m (for $B_0 = 1.55$~T) by performing Biot–Savart simulations of the superconducting solenoid. In this field gradient, displacements on the order of $\sim30$~$\mu$m would be sufficient to cause the experimentally-observed qubit linewidth of 4~kHz (obtained from the measured free induction decay $T_2^* = 80$~$\mu$s with the pulse tube on, data not shown). A similar investigation was carried out by Britton \textit{et al.}, where the same ESR-based spectroscopy technique was used with $\sim 300$ $^9$Be$^+$ ions in a Penning trap to probe the noise resulting from mechanical vibrations (up to 1~kHz) within a homogeneous magnetic field \cite{Britton2015arXiv}.

The third and final possible mechanism for coupling vibrational noise to the qubit coherence is through the creation of voltage noise via triboelectric effects on the cables connected to gate electrodes. This voltage noise would be passed on to the gates thus creating electric field noise at the position of the qubit. As our electron spin qubit is confined by a $^{31}$P donor, the resonant frequency of the qubit would then be affected via Stark shift of the hyperfine interaction and of the electron's $g$-factor \cite{Laucht2015}. As discussed in Section~\ref{sec:CableComparison}, the voltage dividers used at the room-temperature side of the pulsing lines and source-drain lines significantly reduces the amount of voltage noise created at the cable that reaches the device electrodes. From the effective impedance of the `noise source' for the cables used in this setup, we find that voltage noise would be attenuated by at least a factor of 100. Therefore we expect gate noise in our device to be on the order of 100~nV. From the results of an experiment presented in Reference~\onlinecite{Muhonen2014}, we find that a gate noise $>$100~$\mu$V is needed in order to observe a similar amplitude of the noise spectral density as observed in our experiment. This evidence suggests that the more probable mechanisms coupling vibrations from the pulse tube to our qubit is through displacements of the qubits with respect to the applied magnetic field and current noise from the broadband line created by triboelectric effects.

It is interesting to note that the Hahn echo $T_2$ with the pulse tube switched off (Figure~\ref{CoherenceFig}(b), red circles), is 40\% longer than previously measured in the wet fridge (gray diamonds). This could be due to additional filtering of low-frequency thermal radiation on the broadband line achieved by the DC block at the mixing chamber plate, which was not present in the measurements of Reference~\onlinecite{Muhonen2014}. The Hahn echo measurements give encouraging signs that our efforts will lead to better qubit performance, once the effects of the pulse tube noise are mitigated further.

\begin{figure}
	\centering
	\includegraphics[width=\columnwidth]{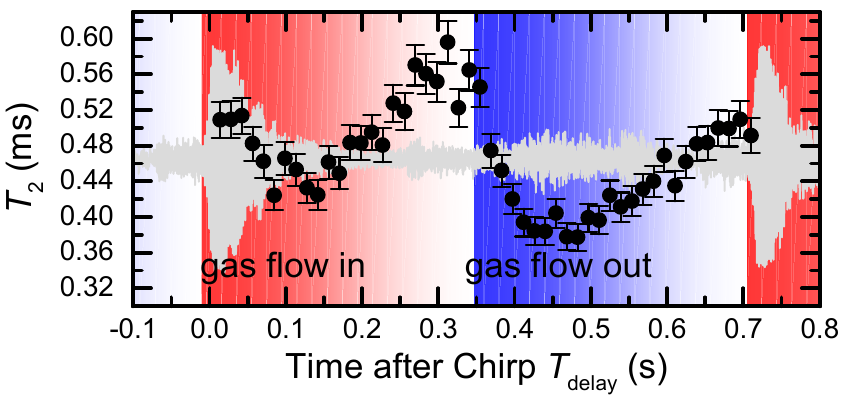}
	\caption{(Color online) Hahn echo coherence time of an electron spin qubit in a BlueFors LD400 cryogen-free dilution refrigerator as a function of time after the pulse tube chirp, $T_\mathrm{delay}$. The black circles correspond to $T_2$ obtained from correlating Hahn echo data to the phase of the pulse tube noise. The gray curve in the background is the corresponding SET readout trace referencing the data to the current noise.}%
	\label{T2vsTchirpFig}
\end{figure}

As a final experiment we test whether or not the electron spin coherence time can be improved by running the Hahn echo experiments synchronized to the chirps of the pulse tube (see Figure~\ref{T2vsTchirpFig}). This is inspired by experiments with atoms in ion traps that are synchronized to the 50~Hz cycle of the power line~\cite{Roos1999}. Of course, this does not work for experiments with duration approaching the period of the pulse tube chirps. The coherence time of the electron spin qubit has been shown to approach 1~s with dynamical decoupling~\cite{Muhonen2014}. Nonetheless, out of interest, we test the effect of synchronization with Hahn echo measurements ($\ll 10$~ms). 

As a proof of principle test, we run the measurement without synchronization and then post-process the data based on the PT noise superimposed on the measurement signal (gray trace in Figure~\ref{T2vsTchirpFig}) to obtain $T_\mathrm{delay}$, the delay between the start of the experiment and the most recent PT noise peak. Hahn echo experiments are run with 12000 shots per wait time $\tau$ to obtain a sufficient number of samples over the range of $T_\mathrm{delay}$. The data is post-processed so that $T_\mathrm{delay}$ is found for each shot. This allows us to bin the decay data into 50 time bins covering the whole PT cycle. On average, this leaves us with 240~shots per $\tau$ and per $T_\mathrm{delay}$, and we have thus created a set of 50 Hahn echo decays for different times after the peak of the PT chirp. We fit the set of Hahn echo decays using global variables for the contrast, offset and exponent of the exponential decay, leaving $T_2$ as the only independent parameter. The $T_2$ values are plotted in Figure~\ref{T2vsTchirpFig} as black circles. The error bar corresponds to the fitting error.

The Hahn echo $T_2$ clearly displays oscillations as a function of $T_\mathrm{delay}$, ranging between $0.38$~ms and $0.60$~ms. Note that the maximum is still well below the Hahn echo $T_2$ measured when the PT was turned off (Figure~\ref{CoherenceFig}(b)). This is not surprising given that some level of vibrations persists through the entire 0.7~s period of the pulse tube cycle (Figure~\ref{SpectroFig}). The $T_2$ data exhibits two clear local minima within a single period, corresponding to the rotary valve connecting the PT to the high and low pressure helium gas lines from the compressor. Recall that this results in helium flowing in and out of the PT as indicated by the red and blue regions in Figure~\ref{T2vsTchirpFig}. One interesting observation is that the minimum $T_2$ time does not coincide with the maximum current noise in the cables. This phase delay may help in determining which of the proposed mechanisms is responsible for coupling the pulse tube noise to the dephasing of the qubit. If we assume that the noise in the semi-rigid cable connected to the antenna and the noise in the flexible cable used for the readout signal are created simultaneously, then the observed delay rules out the conjecture that dephasing is due to current-noise in the antenna. This then supports the conjecture that dephasing is due to movement of the sample inside the inhomogeneous magnetic field, where the delay may come from the inertia of the cold-finger or the solenoid.

\section{Conclusion}

To conclude, we measure electrical noise in our cables that is caused by vibrations originating from the pulse tube cooler. We observe the same spectral comb of peaks in both the electrical signal and in an audio recording of the chirping of the pulse tube. The magnitude of the noise is strongly temperature dependent. We believe that triboelectric effects are the mechanism through which the vibrations couple to the electrical signal. This is consistent with the temperature dependence of the noise amplitude given our model of the cable dielectric undergoing the greatest thermal contraction, leaving room for it to slide against the outer conductor. Flattening a semi-rigid cable or jacketing a flexible cable to reduce movement of the cable dielectric and inner conductor successfully and consistently reduces the pulse tube noise by over an order of magnitude. The two jacketed flexible cables tested perform comparably in terms of both voltage and current noise, both with a significant improvement from the standard semi-rigid cable. While this improvement is sufficient for our purposes, recent work by Mykkanen~\textit{et al.}~\cite{Mykkanen2016arXiv} shows the use of vacuum-insulated cables to efficiently suppress vibration-induced noise.

Aside from introducing current and voltage noise in the measurement and control lines, the pulse tube vibrations also translate into noise that contributes to the decoherence of the spin qubit. Noise spectroscopy measurements on the qubit show a noise spectrum that is qualitatively similar to the pulse tube noise spectrum. It is most likely that this coupling occurs via a combination of mechanical vibrations of the magnet or cold finger and current noise from the broadband line created by triboelectric effects. Future work to address the first issue includes improving the rigidity of the cold finger and further vibrational damping and decoupling. Alternatively, it may be possible to use strong permanent magnets that are bolted to the sample enclosure to replace the superconducting magnet. The second issue can be significantly mitigated by adding a DC block on the broadband line at the device enclosure and by reorienting the sample with respect to the externally applied magnetic field.

\begin{acknowledgments}
We thank David Goldhaber-Gordon for the sample of the graphite-coated cable, Matthew Stead from Resonate Acoustics for the use of the accelerometer, Timothy Duty for the use of the linear micro-stepper driver, and Rob Blaauwgeers, Lieven Vandersypen and Alexander Hamilton for insightful discussions, and David Barber and Rodrigo Ormeno Cortes for technical assistance. We also acknowledge the team involved in the fabrication of the qubit device: Fay E. Hudson, Kohei M. Itoh, David N. Jamieson, Jeffrey C. McCallum and Andrew S. Dzurak. This research was funded by the Australian Research Council Centre of Excellence for Quantum Computation and Communication Technology (project no. CE11E0001027) and the US Army Research Office (W911NF-13-1-0024).
\end{acknowledgments}


\begin{thebibliography}{10}
	
	\bibitem{Tomaru2004}
	T.~Tomaru, T.~Suzuki, T.~Haruyama, T.~Shintomi, A.~Yamamoto, T.~Koyama, and
	R.~Li, ``{V}ibration analysis of cryocoolers,'' {\em Cryogenics}, vol.~44,
	no.~5, pp.~309--317, 2004.
	
	\bibitem{Riabzev2009}
	S.~Riabzev, A.~Veprik, H.~Vilenchik, and N.~Pundak, ``{V}ibration generation in
	a pulse tube refrigerator,'' {\em Cryogenics}, vol.~49, no.~1, pp.~1--6,
	2009.
	
	\bibitem{Baer2014phd}
	S.~Baer, {\em Transport spectroscopy of confined fractional quantum Hall
		systems}.
	\newblock PhD thesis, ETH Zurich, 2014.
	
	\bibitem{Tian2012rsi}
	Y.~Tian, H.~Yu, H.~Deng, G.~Xue, D.~Liu, Y.~Ren, G.~Chen, D.~Zheng, X.~Jing,
	L.~Lu, {\em et~al.}, ``A cryogen-free dilution refrigerator based josephson
	qubit measurement system,'' {\em Review of Scientific Instruments}, vol.~83,
	no.~3, p.~033907, 2012.
	
	\bibitem{Pellicione2013}
	M.~Pelliccione, A.~Sciambi, J.~Bartel, A.~J. Keller, and D.~Goldhaber-Gordon,
	``{D}esign of a scanning gate microscope for mesoscopic electron systems in a
	cryogen-free dilution refrigerator,'' {\em Review of Scientific Instruments},
	vol.~84, no.~3, pp.~--, 2013.
	
	\bibitem{denHaan2014}
	A.~den Haan, G.~Wijts, F.~Galli, O.~Usenko, G.~van Baarle, D.~van~der Zalm, and
	T.~Oosterkamp, ``{A}tomic resolution scanning tunneling microscopy in a
	cryogen free dilution refrigerator at 15 m{K},'' {\em Review of Scientific
		Instruments}, vol.~85, no.~3, p.~035112, 2014.
	
	\bibitem{Ong1987}
	P.~Ong, ``{S}upersensitive electrical measurements and their associated
	techniques,'' {\em European Journal of Physics}, vol.~8, no.~4, p.~280, 1987.
	
	\bibitem{Morello2010}
	A.~Morello, J.~J. Pla, F.~a. Zwanenburg, K.~W. Chan, K.~Y. Tan, H.~Huebl,
	M.~M\"{o}tt\"{o}nen, C.~D. Nugroho, C.~Yang, J.~a. van Donkelaar, A.~D.~C.
	Alves, D.~N. Jamieson, C.~C. Escott, L.~C.~L. Hollenberg, R.~G. Clark, and
	A.~S. Dzurak, ``{Single-shot readout of an electron spin in silicon.},'' {\em
		Nature (London)}, vol.~467, pp.~687--91, Oct. 2010.
	
	\bibitem{Pla2012}
	J.~J. Pla, K.~Y. Tan, J.~P. Dehollain, W.~H. Lim, J.~J.~L. Morton, D.~N.
	Jamieson, A.~S. Dzurak, and A.~Morello, ``{A single-atom electron spin qubit
		in silicon.},'' {\em Nature (London)}, vol.~489, pp.~541--5, Sept. 2012.
	
	\bibitem{Pla2013}
	J.~J. Pla, K.~Y. Tan, J.~P. Dehollain, W.~H. Lim, J.~J.~L. Morton, D.~N.
	Jamieson, A.~S. Dzurak, and A.~Morello, ``{A single-atom electron spin qubit
		in silicon.},'' {\em Nature (London)}, vol.~123, pp.~123--5, Sept. 2013.
	
	\bibitem{Muhonen2014}
	J.~T. Muhonen, J.~P. Dehollain, A.~Laucht, F.~E. Hudson, R.~Kalra,
	T.~Sekiguchi, K.~M. Itoh, D.~N. Jamieson, J.~C. McCallum, A.~S. Dzurak, {\em
		et~al.}, ``{S}toring quantum information for 30 seconds in a nanoelectronic
	device,'' {\em Nature Nanotechnology}, vol.~9, no.~12, pp.~986--991, 2014.
	
	\bibitem{Angus2008}
	S.~Angus, A.~Ferguson, A.~Dzurak, and R.~Clark, ``{A} silicon radio-frequency
	single electron transistor,'' {\em Applied Physics Letters}, vol.~92, no.~11,
	p.~112103, 2008.
	
	\bibitem{Laucht2015}
	A.~Laucht, J.~T. Muhonen, F.~A. Mohiyaddin, R.~Kalra, J.~P. Dehollain,
	S.~Freer, F.~E. Hudson, M.~Veldhorst, R.~Rahman, G.~Klimeck, and et~al.,
	``{E}lectrically controlling single-spin qubits in a continuous microwave
	field,'' {\em Science Advances}, vol.~1, p.~e1500022, Apr 2015.
	
	\bibitem{Alvarez2011}
	G.~A. {\'A}lvarez and D.~Suter, ``Measuring the spectrum of colored noise by
	dynamical decoupling,'' {\em Physical Review Letters}, vol.~107, no.~23,
	p.~230501, 2011.
	
	\bibitem{Hahn1950}
	E.~Hahn, ``{S}pin {E}choes,'' {\em Physical Review}, vol.~80, pp.~580--594, Nov
	1950.
	
	\bibitem{Britton2015arXiv}
	J.~Britton, J.~Bohnet, J.~Bollinger, B.~Sawyer, H.~Uys, and M.~Biercuk,
	``Vibration-induced field fluctuations in a superconducting magnet,'' {\em
		arXiv:1512.00801}, 2015.
	
	\bibitem{Roos1999}
	C.~Roos, T.~Zeiger, H.~Rohde, H.~C. N\"agerl, J.~Eschner, D.~Leibfried,
	F.~Schmidt-Kaler, and R.~Blatt, ``Quantum state engineering on an optical
	transition and decoherence in a paul trap,'' {\em Phys. Rev. Lett.}, vol.~83,
	pp.~4713--4716, Dec 1999.
	
	\bibitem{Mykkanen2016arXiv}
	E.~Mykk{\"a}nen, J.~Lehtinen, A.~Kemppinen, C.~Krause, D.~Drung,
	J.~Nissil{\"a}, and A.~Manninen, ``Reducing current noise in cryogenic
	experiments by vacuum-insulated cables,'' {\em arXiv:1604.03903}, 2016.
	
\end{thebibliography}

\end{document}